# Superconductivity Above 80 K in Polyhydrides of Hafnium

C.L. Zhang[a1,2], X. He [a1,2,3], Z.W. Li [1,2], S.J. Zhang[1], B. S. Min[1,2], J. Zhang[1,2], K. Lu[1,2], J.F. Zhao[1,2], L.C. Shi[1,2], Y. Peng[1,2], X.C. Wang*[1,2], S.M. Feng[1], R.C. Yu[1,2], L.H. Wang[4], V. B. Prakapenka[5], S. Chariton[5], H. Z. Liu[6], C. Q. Jin*[1,2,3]

[1] Beijing National Laboratory for Condensed Matter Physics, Institute of Physics, Chinese Academy of Sciences, Beijing 100190, China
[2] School of Physical Sciences, University of Chinese Academy of Sciences, Beijing 100190, China
[3] Songshan Lake Materials Laboratory, Dongguan 523808, China
[4] Shanghai Advanced Research in Physical Sciences, Shanghai 201203, China
[5] Center for Advanced Radiations Sources, University of Chicago, Chicago, Illinois 60637, USA
[6] Center for High Pressure Science & Technology Advanced Research, Beijing 100094, China

**Abstract**

Studies on polyhydrides are attracting growing attentions recently due to their potential high temperature superconductivity (SC). We here report the discovery of SC in hafnium polyhydrides at high pressures. The hafnium superhydrides are synthesized at high pressure and high temperature conditions using diamond anvil cell in combination with *in-situ* high pressure laser heating technique. The SC was investigated by *in-situ* high pressure resistance measurements in applied magnetic fields. A superconducting transition with onset $T_c$ ~83 K was observed at 243 GPa. The upper critical field $\mu_0 H_{c2}(0)$ was estimated to be 24 Tesla by GL theory and the consequent superconducting coherent length $\xi$ to be ~37 Å. Our results suggest that the superconducting phase is from $C2/m$-HfH$_{14}$. This is the first 5$d$ transition metal polyhydride superconductor with $T_c$ above the liquid nitrogen temperature.

**Introduction**

The research of superconductivity (SC) in polyhydrides has received intensively attentions since the high critical transition temperature $T_c$ almost approaches room temperature[1-3]. The high $T_c$ SC is considered to arise from the metallized hydrogen framework on the basis of Bardeen-Cooper-Schrieffer theory. In the regard the electrons from hydrogen have considerable contribution to the density of state near the Fermi surface. Since hydrogen is the lightest element and thus possesses high Debye temperature, polyhydrides would be expected to exhibit high $T_c$ SC if the hydrogen framework metallization occurs. Several polyhydrides have been theoretically predicted and experimentally discovered to possess high $T_c$ SC under high pressure[4-20]. The $SH_3$ is the first superhydride that was reported to exhibit SC with $T_c$ exceeding 200 K[4, 5]. Then a series of metal polyhydrides superconductors are experimentally discovered such as $LaH_{10}$, $YH_9$ & $CaH_6$ *etc* [6-20]. On the basis of discovered binary superhydride superconductors, it was found that most of the elemental metals are located on the boundary of *s* and *d* blocks in the elemental period table, that is within the IIA and IIIB groups. It is suggested that pressure induced *s-d* electron transfer due to the small energy difference between (n+1)-*s* and n-*d* orbitals should enhance the electron-phonon interaction and $T_c$[2, 19]. Hence great efforts have made to search for new high $T_c$ superhydride superconductors formed by the metal in IVB group[21-25]. It is generally believed that *d* electrons are harmful to the conventional BCS superconductors since local spin will weaken the Cooper pair. This seems to be obvious if one take into account the fact that the early polyhydrides such as $SH_3$, rare earth hydrides and calcium hydrides are all *d* electron free. However, more recently, zirconium polyhydrides was experimentally reported to be superconducting with $T_c$ ~70 K at high pressures that is so far the only 4*d* transition metal hydrides[18]. Hafnium element itself is a superconductor with $T_c$ ~0.1 K at ambient[26] that can be increased to ~8 K at 60 GPa[27]. Although hafnium hydrides was reported to be superconductor[22], the $T_c$ ~5 K was found to be comparable to high pressure Hf element that shows superconductivity up to ~8 K at 60 GPa. Here, we report the synthesis of hafnium superhydride and the discovery of SC with $T_c$ exceeding 83 K at 243 GPa. It is the first 5*d* transition metal superhydride with $T_c$ exceeding the liquid nitrogen temperature.

**Methods**

For the synthesis of hafnium superhydride and *in-situ* high pressure resistivity measurements, diamond anvil cells (DAC) made of nonmagnetic Cu-Be alloys was used to generate ultrahigh pressure more than 2 Mbar. The culet diameter of the diamond anvils used in DAC is 50 μm beveled to diameter of 300 μm. The pre-pressed T301 stainless steel with a hole of 300 μm in diameter was used as the gasket to hold the sample under high pressure. Aluminum oxide mixed with epoxy resin was filled into the hole, pre-pressed and drilled to form a high pressure chamber with 40 μm in diameter. Then, ammonia borane (AB) was filled into the chamber as the hydrogen source and the pressure transmitting medium as well. The Pt foils (0.5 μm in thickness) were deposited on the surface of the top anvil culet to form four inner electrodes, which can be well attached by outside electrodes of gold wire. High purity hafnium metal (99.9%) with the size of 20 μm×20 μm and 1 μm in thickness was stacked on the inner electrodes before the DAC was clamped. To avoid moisture or oxidation, the sample loading was carried out in a glove box filled with Ar gas with 1 ppm less trace water and oxygen. The details are referred to the ATHENA procedure reported in ref.[28].

After applying high pressure, the mixture of Hf and AB was heated at 2000 K for several minutes by laser with a focused spot size 5 μm in diameter. A YAG laser in a continuous mode with 1064 nm wave length was adopted for the laser heating. The temperature was determined by fitting the black body irradiation spectra. The pressure was determined using the Raman peak of diamond. The sample quenched from high temperature without releasing pressure was put into a MagLab system for the electric conductivity measurements. A Van der Pauw method was employed with an applied electric current of 1 mA. The MagLab system can provide synergetic extreme environments with temperatures from 300 K to 1.5 K and a magnetic field up to 9T[29, 30].

*In-situ* high-pressure x-ray diffraction (XRD) patterns were collected at room temperature at 13-IDD of Advanced Photon Source at the Argonne National Laboratory with the wavelength λ = 0.3344 Å. The x-ray was focused down to a spot of ~3 μm in diameter. A symmetric diamond anvil cell was used for the XRD experiments. A prepressed rhenium gasket with a hole with diameter of 25 μm was used as the high pressure chamber. Samples were loaded into the pressure chamber without aluminum oxide insulating layer. After applying high pressure the sample was laser heating to synthesize hafnium superhydride. The sample pressure in the diffraction experiments was calibrated by the equation of state for rhenium and internal pressure marker Pt. The XRD images are converted to one dimensional diffraction data with Dioptas[31].

**Results and discussions**

Hafnium superhydrides sample was synthesized at the synergetic conditions of above 200 GPa and approximately 2000 K. Fig. 1(a) shows the image of the experimental assembly of DAC. The yellow circle represents the culet of the diamond and the red square shows the shape of sample. The four green areas shows the four inner Pt electrodes which is contacted well with the sample. The temperature dependence of resistance $R(T)$ for sample A measured at 243 GPa and sample B at 212 GPa are shown in Fig. 1.(b). For sample A, the curves measured in both cooling and warming processes are almost overlapped, indicating little temperature hysteresis for the measurement. The resistance drops rapidly at ~83 K and reaches zero at ~78 K, implying an occurrence of superconducting transition. The inset of Fig. 1.(b) is the temperature derivative of resistance (d$R$/d$T$) for the warming process. The onset superconducting temperature $T_c^{onset}$ is determined with the right upturn of d$R$/d$T$ curve to be ~83 K. From the d$R$/d$T$ curve, the $T_c^{zero}$ where the resistance reaches zero can be determined as well by the left upturn to be ~78 K. For sample B, the onset $T_c$ at 212 GPa is changed little compared with sample A. The lower normal resistant relative to sample A arises from the difference of sample shape in the high pressure chamber. It seems that pressure has little effect on $T_c$ in the range of 212 GPa - 243 GPa. For hafnium superhydrides, some first principle calculations have predicted the structure and SC under high pressure. The $T_c$ was proposed to be 43 K at 100 GPa for $Hf_3H_{13}$[2], 110 K at 200 GPa for $HfH_9$[24], 234 K at 250 GPa for $HfH_{10}$[23], 76 K at 300 GPa for $HfH_{14}$[2] and 132 K at 600 GPa for $HfH_6$[25]. In the pressure range of 200 GPa - 300 GPa, the observed $T_c$ here is much lower than that predicted for $HfH_9$ and $HfH_{10}$, but very close to that of $C2/m$-$HfH_{14}$. Therefore, it is suggested the observed SC should arise from $C2/m$-$HfH_{14}$, which will be further confirmed by the following XRD experiments. In addition to zirconium superhydride ($T_c$ ~71 K)[18], hafnium superhydride is another transition metal superhydride superconductor with $T_c$ near the liquid nitrogen temperature. For the IIA and IIIB group metals of Ca, Y and La, pressure induced $s$-$d$ electrons transfer is suggested to play an important role for the high temperature SC of their metal superhydrides[2, 19]. The $s$-$d$ electrons transfer under high pressure would increase the structure instability and tend to enhance the electron-phonon coupling and thus lead to high $T_c$. Therefore, the vicinity of the region of IIA and IIIB groups should be an important reason for the high $T_c$ SC observed in hafnium superhydride.

To further investigate the property of the SC in the synthesized hafnium superhydride sample, the *in-situ* high pressure resistance measured under high magnetic field has been performed, as shown in Fig. 2(a). When increasing magnetic field, the superconducting transition is gradually suppressed and the transition width increases from ~5 K at zero field to ~10 K at 5 T. The critical field $H_{c2}$ versus $T_c$ is plotted in Fig. 2(b), which presents a straight line. Here, the $T_c$ values were determined with the temperature at 95% and 50% of normal resistance for $T_c^{95\%}$ and $T_c^{50\%}$, respectively and the criteria of zero resistance for $T_c^{zero}$. After the linear fitting of the $H_{c2}(T)$ data, the slope of $dH_c/dT$ can be obtained to be = -0.55 T/K, -0.37 T/K and -0.30 T/K for the criteria of $T_c^{95\%}$, $T_c^{50\%}$ and $T_c^{zero}$, respectively. The upper critical magnetic field at zero temperature of $H_{c2}(0)$ can be estimated by the Werthamer-Helfand-Hohenberg (WHH) formula of $\mu_0 H_{c2}(T) = -0.69 \times dH_{c2}/dT\,|_{T_c} \times T_c$, which was obtained to be 31 T for the criteria of $T_c^{95\%}$, 20 T for $T_c^{50\%}$ and 16 T for $T_c^{zero}$. The slope of $dH_c/dT$ is almost half of that for ZrH$_n$(-0.99 T/K with $T_c^{onset}$ =70 K)[18] and one fourth of that for CaH$_n$ with the slope of -1.97 T/K and $T_c$ ~210 K[19], which suggests that the magnetic vortex pinning force in hafnium superhydride is very weak. In addition, $\mu_0H_{c2}(0)$ can also be estimated by the Ginzburg Landau (GL) formula of $\mu_0H_{c2}(T) = \mu_0H_{c2}(0)(1-(T/T_c)^2)$. As shown in Fig. 2(b) the fitting of the $\mu_0H_{c2}(T)$ by GL formula gives a value of $\mu_0H_{c2}(0)$ ranging from 13 T to 24 T using the criteria of $T_c^{95\%}$, $T_c^{50\%}$ and $T_c^{zero}$, respectively. The GL coherent length $\xi$ can be estimated to be ~37 Å by the equation of $\mu_0H_{c2}(0)= \Phi_0/2\pi\xi^2$, where $\Phi_0= 2.067\times10^{-15}$ Web is the magnetic flux quantum and $\mu_0H_{c2}(0)= 24$ T is used.

To investigate the superconducting phase, high pressure XRD experiments were carried out. Fig. 3 shows the XRD patterns collected under 206 GPa and its refinement, from which it can be seen that majority of the diffraction peaks are contributed by $C2/m$-HfH$_{14}$. The refinement generates the lattice parameters of $a$=4.72 Å, $b$=3.33 Å, $c$=4.70 Å and $\beta$=90.27°. The corresponding unit cell volume at 206 GPa are ~22% larger than that of $C2/m$-HfH$_{14}$ predicted at 300 GPa[2]. There are some unknown weak diffraction peaks in the XRD pattern, which are marked by the arrows at 2 theta at 7.73°, 8.52° and 10.17°, respectively. Since these unknown peaks cannot be assigned as the gasket Re metal, pressure marker Pt or boron nitride generated from the decomposition of AB, it is suggested they should arise from other hafnium hydrides with different hydrogen content. Therefore, combining with the superconductivity studies and the structure analysis, we proposed that the $C2/m$-HfH$_{14}$ is responsible for the observed SC with $T_c$ ~83 K at this pressure region. The inset of Fig. 3 shows the schematic view of $C2/m$-HfH$_{14}$ crystal structure, wherein most of the hydrogen atoms are bonded with the bond length less than 0.9 Å. That is why the $T_c$ of HfH$_{14}$ is lower than that for CaH$_6$ (H-H length ~1.24 Å) and LaH$_{10}$ (H-H length ~1.1 Å) etc. although its hydrogen content is higher.

In summary, a new hafnium superhydride superconductor was synthesized under high pressure and high temperature conditions. The superconducting transition was observed with $T_c$ ~ 83 K at 243 GPa. Combining with the SC and structure studies, it is suggested that the superconducting phase should be $C2/m$-HfH$_{14}$ phase.

**Acknowledgements**
The work was supported by NSF, MOST & CAS of China through research projects.

**Credit author statement**

C.L. Zhang: Investigation；X. He：Methodology；Z.W. Li：Methodology；S.J. Zhang：Methodology；B. S. Min: Investigation：J. Zhang：Investigation；K. Lu：Investigation；J.F.Zhao: Investigation ；L.C.Shi: Investigation ；Y. Peng: Investigation；X.C. Wang: Investigation, Writing, Editing.；S.M. Feng[1], R.C. Yu: Investigation；L.H. Wang：Methodology；V. B. Prakapenka：Methodology；Methodology；S. Chariton：Methodology；H. Z. Liu：Methodology；C. Q. Jin: Conceptualization, Supervision, Writing.

**Declaration of competing interest**

The authors declare no competing financial interest.

**References**

[1] N. W. Ashcroft, "Hydrogen dominant metallic alloys: High temperature superconductors?". Phys. Rev. Lett. 92, 187002 (2004).

[2] D. V. Semenok, I. A. Kruglov, I. A. Savkin, A. G. Kvashnin, A. R. Oganov, "On Distribution of Superconductivity in Metal Hydrides". Curr Opin Solid St M 24, 100808 (2020).

[3] J. A. Xu, Z. W. Zhu, "Metallic Hydrogen". Physics 6, 296 (1977).

[4] D. F. Duan, Y. X. Liu, F. B. Tian, D. Li, X. L. Huang, Z. L. Zhao, H. Y. Yu, B. B. Liu, W. J. Tian, T. Cui, "Pressure induced metallization of dense (H2S)(2)H2 with high Tc superconductivity". Sci. Rep. 4, 6968 (2014).

[5] A. P. Drozdov, M. I. Eremets, I. A. Troyan, V. Ksenofontov, S. I. Shylin, "Conventional superconductivity at 203 kelvin at high pressures in the sulfur hydride system". Nature 525, 73 (2015).

[6] H. Wang, J. S. Tse, K. Tanaka, T. Iitaka, Y. M. Ma, "Superconductive sodalite-like clathrate calcium hydride at high pressures". PNAS 109, 6463 (2012).

[7] F. Peng, Y. Sun, C. J. Pickard, R. J. Needs, Q. Wu, Y. M. Ma, "Hydrogen Clathrate Structures in Rare Earth Hydrides at High Pressures: Possible Route to Room-Temperature Superconductivity". Phys. Rev. Lett. 119, 107001 (2017).

[8] H. Y. Liu, I. I. Naumov, R. Hoffmann, N. W. Ashcroft, R. J. Hemley, "Potential high Tc superconducting lanthanum and yttrium hydrides at high pressure". PNAS 114, 6990 (2017).

[9] P. P. Kong, V. S. Minkov, M. A. Kuzovnikov, A. P. Drozdov, S. P. Besedin, S. Mozaffari, L. Balicas, F. F. Balakirev, V. B. Prakapenka, S. Chariton, D. A. Knyazev, E. Greenberg, M. I. Eremets, "Superconductivity up to 243 K in the yttrium-hydrogen system under high pressure". Nat. Commun. 12, 5075 (2021).

[10] J. A. Flores-Livas, L. Boeri, A. Sanna, G. Profeta, R. Arita, M. Eremets, "A perspective on conventional high-temperature superconductors at high pressure: Methods and materials". Phys. Rep. 856, 1 (2020).

[11] Z. M. Geballe, H. Y. Liu, A. K. Mishra, M. Ahart, M. Somayazulu, Y. Meng, M. Baldini, R. J. Hemley, "Synthesis and Stability of Lanthanum Superhydrides". Angew. Chem. Int. Edit 57, 688 (2018).

[12] A. P. Drozdov, P. P. Kong, V. S. Minkov, S. P. Besedin, M. A. Kuzovnikov, S. Mozaffari, L. Balicas, F. F. Balakirev, D. E. Graf, V. B. Prakapenka, E. Greenberg, D. A. Knyazev, M. Tkacz, M. I. Eremets, "Superconductivity at 250 K in lanthanum hydride under high pressures". Nature 569, 528 (2019).

[13] M. Somayazulu, M. Ahart, A. K. Mishra, Z. M. Geballe, M. Baldini, Y. Meng, V. V. Struzhkin, R. J. Hemley, "Evidence for Superconductivity above 260 K in Lanthanum Superhydride at Megabar Pressures". Phys. Rev. Lett. 122, 27001 (2019).

[14] F. Hong, L. X. Yang, P. F. Shan, P. T. Yang, Z. Y. Liu, J. P. Sun, Y. Y. Yin, X. H. Yu, J. G. Cheng, Z. X. Zhao, "Superconductivity of Lanthanum Superhydride Investigated Using the Standard Four-Probe Configuration under High Pressures". Chin. Phys. Lett. 37, 107401 (2020).

[15] E. Snider, N. Dasenbrock-Gammon, R. Mcbride, X. Wang, N. Meyers, K. V. Lawler, E. Zurek, A. Salamat, R. P. Dias, "Synthesis of Yttrium Superhydride Superconductor with a Transition Temperature up to 262 K by Catalytic Hydrogenation at High Pressures". Phys. Rev. Lett. 126, 117003 (2021).

[16] F. Hong, P. F. Shan, L. X. Yang, B. B. Yue, P. T. Yang, Z. Y. Liu, J. P. Sun, J. H. Dai, H. Yu, Y. Y. Yin, X. H. Yu, J. G. Cheng, Z. X. Zhao, Mater Today Phys 22, 100596(2022)

[17] W. Chen, D. V. Semenok, A. G. Kvashnin, X. Huang, I. A. Kruglov, M. Galasso, H. Song, D. Duan, A. F. Goncharov, V. B. Prakapenka, A. R. Oganov, T. Cui, "Synthesis of molecular metallic barium superhydride: pseudocubic BaH12". Nat. Commun. 12, 273 (2021).

[18] C. L. Zhang, X. He, Z. W. Li, S. J. Zhang, S. M. Feng, X. C. Wang, R. C. Yu, C. Q. Jin, Sci Bull. 67, 907 (2022)

[19] Z. Li, X. He, C. L. Zhang, X. C. Wang, S. J. Zhang, Y. T. Jia, S. M. Feng, K. Lu, J. F. Zhao, J. Zhang, B. S. Min, Y. W. Long, R. C. Yu, L. H. Wang, M. Y. Ye, Z. S. Zhang, V. Prakapenka, S. Chariton, P. A. Ginsberg, J. Bass, S. H. Yuan, H. Z. Liu, C. Q. Jin, Nat. Commun. 13, 2863(2022)

[20] L. Ma, K. Wang, Y. Xie, X. Yang, Y. Wang, M. Zhou, H. Liu, X. Yu, Y. Zhao, H. Wang, G. Liu, Y. Ma, Phys. Rev. Lett. 128, 167001(2022)

[21] D. V. Semenok, A. G. Kvashnin, A. G. Ivanova, V. Svitlyk, V. Y. Fominski, A. V. Sadakov, O. A. Sobolevskiy, V. M. Pudalov, I. A. Troyan, A. R. Oganov, Materials Today 33, 36(2020)

[22] M. A. Kuzovnikov, M. Tkacz, The Journal of Physical Chemistry C 123, 30059 (2019)

[23] H. Xie, Y. S. Yao, X. L. Feng, D. F. Duan, H. Song, Z. H. Zhang, S. Q. Jiang, S. a. T. Redfern, V. Z. Kresin, C. J. Pickard, T. Cui, Phys. Rev. Lett. 125, 217001(2020)

[24] K. Gao, W. Cui, J. Chen, Q. Wang, J. Hao, J. Shi, C. Liu, S. Botti, M. a. L. Marques, Y. Li, Phys. Rev. B 104, 214511 (2021)

[25] P. Tsuppayakorn Aek, N. Phaisangittisakul, R. Ahuja, T. Bovornratanaraks, Sci. Rep. 11, 16403(2021)

[26] R. A. HEIN, Physical Review 102, 1511(1956)

[27] I. O. Bashkin, M. V. Nefedova, V. G. Tissen, E. G. Ponyatovsky, Jetp Letters 80, 655 (2004)

[28] Y. T. Jia, X. He, S. M. Feng, S. J. Zhang, C. L. Zhang, C. W. Ren, X. C. Wang, C.

Q. Jin, "A Combinatory Package for Diamond Anvil Cell Experiments". Crystals 10, 1116 (2020).

[29] J. L. Zhang, S. J. Zhang, H. M. Weng, W. Zhang, L. X. Yang, Q. Q. Liu, S. M. Feng, X. C. Wang, R. C. Yu, L. Z. Cao, L. Wang, W. G. Yang, H. Z. Liu, W. Y. Zhao, S. C. Zhang, X. Dai, Z. Fang, C. Q. Jin, "Pressure-induced superconductivity in topological parent compound Bi2Te3", PNAS 108, 24 (2011).

[30] S. J. Zhang, X. C. Wang, R. Sammynaiken, J. S. Tse, L. X. Yang, Z. Li, Q. Q. Liu, S. Desgreniers, Y. Yao, H. Z. Liu, C. Q. Jin, "Effect of pressure on the iron arsenide superconductor LixFeAs (x=0.8,1.0,1.1)". Phys. Rev. B 80, 14506 (2009).

[31] C. Prescher, V. B. Prakapenka, "DIOPTAS: a program for reduction of two-dimensional X-ray diffraction data and data exploration". High Press. Res. 35, 223 (2015).

**Figure Captions**

(**All figures are in color form in the printed version**)

Fig. 1. (a) The image of the high pressure assembly of DAC. (b) Temperature dependence of resistance $R(T)$ for sample A measured at 243 GPa and sample B at 212 GPa. The inset is the temperature derivative of resistance (d$R$/dT), where the $T_c^{onset}$ and $T_c^{zero}$ are determined to be ~83 K and 78 K, respectively.

Fig. 2. (a) Resistance measured under different magnetic field and under pressure of 243 GPa. (b) The upper critical magnetic field $H_{c2}$(T) versus temperature with the criteria of $T_c^{95\%}$, $T_c^{50\%}$ and $T_c^{zero}$, respectively. The solid lines are the fitting via GL theory; The dashed straight lines are the linear fitting. The stars show the estimated $H_{c2}(0)$ values by WHH method.

Fig. 3 *In-situ* high pressure XRD and its refinement. The inset shows the schematic view of $C2/m$-HfH$_{14}$ structure.

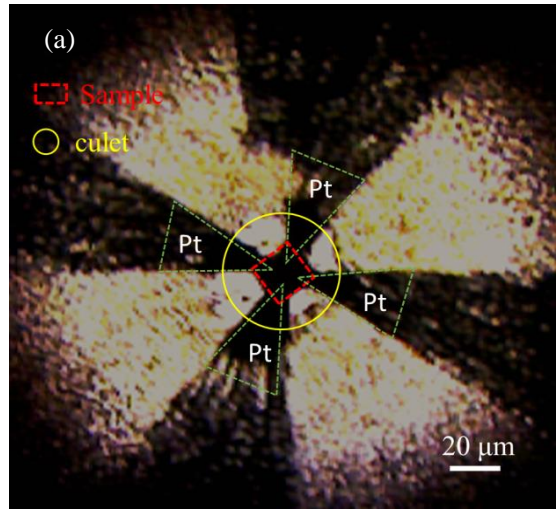

Fig. 1. (a) The image of the high pressure assembly of diamond anvil cell.

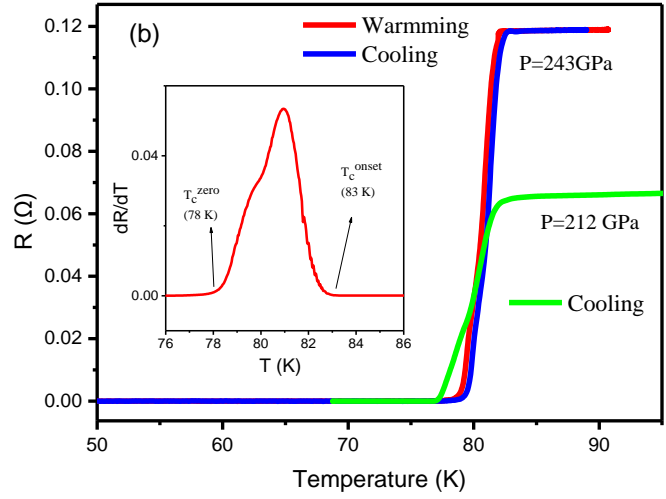

Fig. 1. (b) Temperature dependence of resistance $R(T)$ for sample A measured at 243 GPa and sample B at 212 GPa. The inset is the temperature derivative of resistance ($dR/dT$), where the $T_c^{onset}$ and $T_c^{onset}$ are determined to be ~83 K and 78 K, respectively.

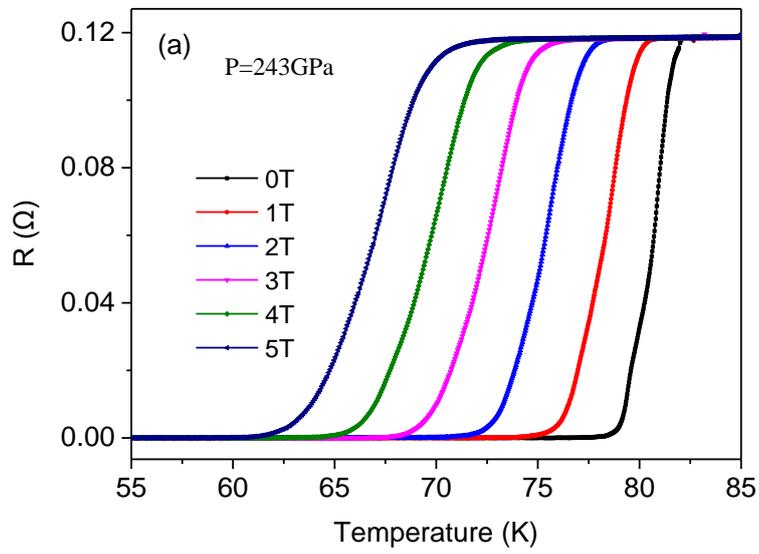

Fig. 2. (a) Resistance measured under different magnetic field and under pressure of 243 GPa.

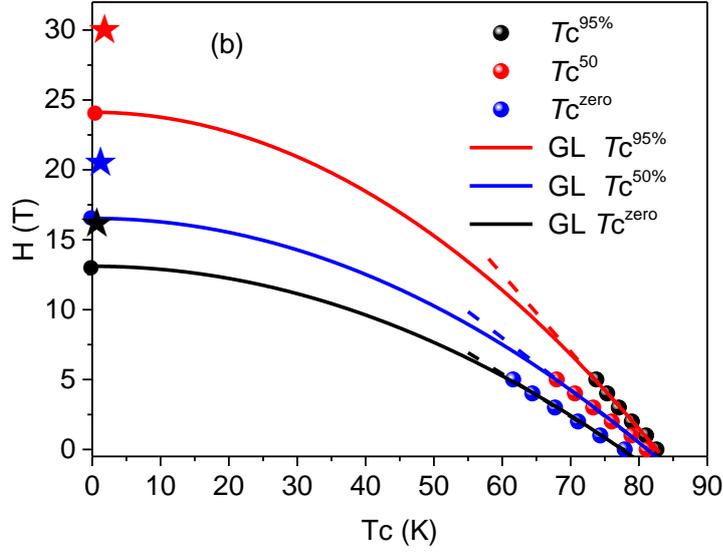

Fig. 2. (b) The upper critical magnetic field $H_{c2}(T)$ versus temperature with the criteria of $T_c^{95\%}$, $T_c^{50\%}$ and $T_c^{zero}$, respectively. The solid lines are the fitting via GL theory; The dashed straight lines are the linear fitting. The stars show the estimated $H_{c2}(0)$ values by WHH method.

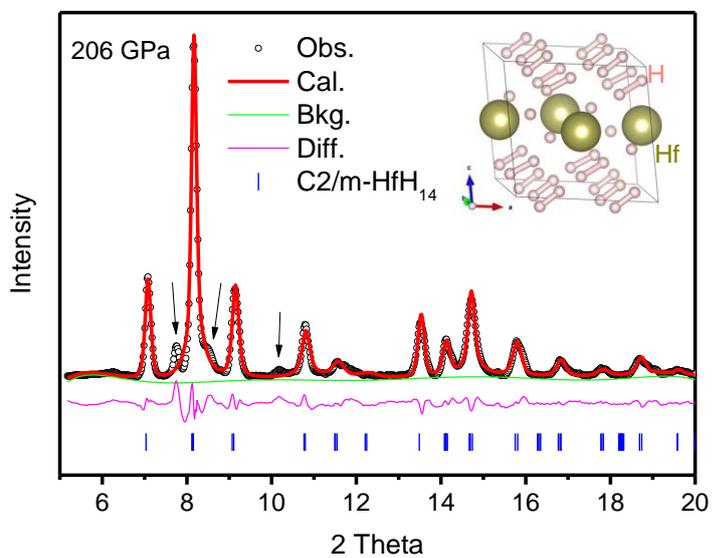

Fig. 3 *In-situ* high pressure XRD and its refinement. The inset shows the schematic view of $C2/m$-HfH$_{14}$ structure.